# The 2 mrad crossing-angle ILC interaction region and extraction line

**R. Appleby[1]**
*Cockcroft Institute and the University of Manchester, Oxford Road, Manchester M139PL, England*

**D. Angal-Kalinin**
*CCLRC, ASTeC, Cockcroft Institute, Daresbury Laboratory,
Chilton, DIDCOT, OX11 0QX, England*

**O. Dadoun, P. Bambade**
*Laboratoire de l'Accélérateur Linéaire,
IN2P3-CNRS et Université de Paris-Sud 11- Bât. 200, BP 34, 91898 Orsay cedex, France*

**B. Parker**
*Brookhaven National Lab., P.O. Box 5000, Upton, NY 11973-5000*

**L. Keller, K. Moffeit, Y. Nosochkov, A. Seryi, C. Spencer**
*Stanford Linear Accelerator Center, 2575 Sand Hill Road, Menlo Park, CA 94025, USA*

**J. Carter**
*John Adams Institute and Royal Holloway, University of London, England*

**O. Napoly**
*CEA/DSM/DAPNIA-Saclay, 91191 Gif-sur-Yvette, France*

*Abstract*

A complete optics design for the 2mrad crossing angle interaction region and extraction line was presented at Snowmass 2005. Since this time, the design task force has been working on developing and improving the performance of the extraction line. The work has focused on optimising the final doublet parameters and on reducing the power losses resulting from the disrupted beam transport. In this paper, the most recent status of the 2mrad layout and the corresponding performance are presented.

[1] #r.b.appleby@dl.ac.uk. Work supported in part by the Commission of European Communities under the 6th Framework Programme Structuring the European Research Area, contract number RIDS-011899

## INTRODUCTION

The International Linear Collider (ILC) baseline design consists of two interaction regions: one with a large crossing angle and one with a small crossing angle. The small crossing angle layout [1] is favoured by the physics case, although it presents considerable design and technological challenges. The first complete layout was presented at Snowmass 2005 [2], and is documented in [3]. We refer to this design as the baseline layout. However, this design is optimised for 1 TeV centre of mass energy (CM) and there was a scope to increase the final quadrupole gradient further. It was previously assumed that exactly the same optics can be used at 500 GeV CM by scaling down the field in all magnets. However, the use of a long final doublet with low field was shown to be not optimum at 500 GeV CM and causing unsatisfactory beam power losses for some of the ILC parameter sets [4]. In this paper, we describe the baseline layout, consider current and proposed magnet technologies to redesign the interaction region, and present the first design, including extraction line optics, optimised specifically for the 500 GeV CM machine. Since the ILC will operate at 500 GeV CM for first few years, it seems natural to have a better design with good performance at 500 GeV to start with. The design needs to work at 1 TeV and the final doublet can be replaced during the upgrade to 1 TeV.

This paper describes the optimised doublet at 500 GeV and shows the performance improvement compared to the Snowmass design. The extraction line is rematched with this new doublet and the downstream line includes the beam diagnostics as in the baseline version.

## PERFORMANCE OF SNOWMASS FINAL DOUBLET REGION

The baseline layout was first presented at Snowmass 2005 for l*=4.5m [2]. The design includes beamstrahlung tail collimation and energy/polarisation diagnostics. The extraction line has been designed for 1 TeV CM, with the 500 GeV CM optics obtained by scaling down the fields. The final doublet region is designed using a superconducting large bore final quadrupole magnet (denoted QD0), superconducting large bore sextupoles (denoted by SD0 and SF1) and a warm pocket field quadrupole (denoted QF1) magnet. The choice of superconductor for QD0 is NbTi, with a maximum pole tip field of 5.6T. The outgoing beam passes off-axis through the incoming beam QD0 and sextupole magnets, by virtue of the crossing angle, and receives a beam-separating horizontal kick.

The beam transport properties have been extensively studied since Snowmass 2005, and presented in [3]. Table 1 shows the sum of the disrupted charged beam and radiative Bhabha (RB) losses into the final doublet magnets for the low power beam parameter set [4]. The latter are produced during the intense beam-beam interaction during collision. The Low Power parameter set has a large beamstrahlung parameter, which makes the beam extraction challenging. Similarly, the very high beamstrahlung parameter of the High Luminosity parameter set makes the beam extraction unfeasible; hence we do not consider it in this work. We also calculate the power losses for the case of a vertical offset at the interaction point. The offset is chosen to maximise the beamstrahlung radiation during collisions. These power losses are high, and can be shown to quench the superconducting magnets for several parameter sets [6], and hence the baseline layout does not work in required regions of the ILC parameter space.

## REDESIGNED FINAL DOUBLETS

We have used recent and proposed advances in superconducting magnet technology to optimise the final oublet region [5]. NbTi magnets can, with current coil designs, reach a pole tip field of 6.3T [5]. This includes the field-reducing effects of external solenoid fields and an additional safety margin. We shall use the NbTi based technology and call it a basic technology in this paper, along with superconducting sextupole magnets with a pole tip field of up to 4 T, to redesign the final doublet region for the 500 GeV and the 1 TeV layouts. Note that this is the first specific optimisation for 500 GeV. We also use $Nb_3Sn$ superconducting magnet technology for QD0, with a maximum pole tip field of 8.75T [5]. Hence we propose three new doublet layouts

- NbTi layout for the 500 GeV machine
- NbTi layout for the 1 TeV machine
- $Nb_3Sn$ layout for the 1 TeV machine

These magnets are currently under R & D, but on the timescale of the ILC they will be ready and can be thought of as representative of magnet advances over this period.

| | LOCATION | | | |
|---|---|---|---|---|
| Beam | QD0 | SD0 | QF1 | SF1 |
| Low power | 7.6 W | 0 W | 0 W | 469.8 W |
| Low power (dy=120nm) | 7.8W | 0 W | 0 W | 188.1 W |
| Low power RB | 0.4W | 0W | 0.01 W | 0.1 W |

Table 1: The disrupted charged beam and radiative Bhabha (RB) losses into the final doublet magnets for the baseline 500 GeV final doublet. We only show the low power parameter set losses, and dy denotes a vertically offset beam at the collision point.

| Magnet | Length | Strength | Radial aperture |
|---|---|---|---|
| QD0 | 1.2344m | -0.194 $m^{-1}$ | 39mm |
| SD0 | 2.5m | 1.1166 $m^{-2}$ | 76mm |
| QF1 | 1.0m | 0.08146 $m^{-1}$ | 15mm |
| SF1 | 2.5m | -0.2731 $m^{-2}$ | 151mm |

Table 2: The magnet parameters for the optimised 500 GeV NbTi final doublet.

In this paper, we discuss the new final doublet design, focusing specifically on the NbTi 500 GeV machine layout. The sextupoles shall be based on baseline technology, and can be re-optimised. The final doublet is designed to take account of the total power loss from disrupted charged beam and radiative Bhabha transport. The parameter spaces of the magnets are scanned to minimise this figure of merit. The details of the procedure can be found in [6], and the resulting magnet parameters for the new NbTi 500 GeV layout can be seen in table 2. Table 3 shows the combined power losses into the magnets for the NbTi 500 GeV new final double layout. Again, we consider the Low Power Parameter set and the case of a vertical offset at the interaction point. The reduction in losses, compared to table 1, can be seen, demonstrating the improved beam transport properties of the new design. The optimisation procedure leads to an increase in the aperture of of QD0 from 35mm to 39mm. However, as was shown in [6], the minimum obtained is shallow enough that 35mm could also be acceptable. These losses could be further reduced, and the magnets shortened, by using $Nb_3Sn$ for the 500 GeV layout. We also have a full set of magnet parameters for the other two cases: NbTi and $Nb_3Sn$ at 1 TeV.

| | LOCATION | | | |
|---|---|---|---|---|
| Beam | QD0 | SD0 | QF1 | SF1 |
| Low power | 0 W | 0 W | 0 W | 0 W |
| Low power (dy=120nm) | 0 W | 0 W | 0 W | 0 W |
| Low power RB | 0.05 W | 0.1 W | 0.13 W | 0.03 W |

Table 3: The disrupted charged beam and radiative Bhabha (RB) losses into the final doublet magnets for the optimised 500 GeV NbTi final doublet. We only show the low power parameter set losses, and dy denotes a vertically offset beam at the collision point.

The localised peak power deposition into the coils of the superconducting magnets can be reduced using Tungsten liner on the particle deposition hot spot. The use of such a liner was considered in [6], and it was found that the resulting localised power depositions could be made less than the quench limit of the magnet.

The specific requirements of the $Nb_3Sn$ magnets for the International Linear Collider are also under study. For example, flux jumping may occur in regions of the magnetic coils where the field is low and this effect will increase when the magnet is used for low energy beams [7]. The impact of such flux jumping on the ILC beam needs to be assessed, especially when the energy will be changed by a factor of 5 (for 500 GeV CM) during the calibration and operation.

## DOWNSTREAM INTEGRATION

The optimised final doublet has been integrated into the baseline downstream optics, and the extraction line specifically rematched for the 500 GeV CM machine. The other two new doublet layouts need to be integrated into the downstream optics but we focus on the 500 GeV machine in this paper. The extraction line optics was constrained to provide a beam suitable for polarimetry at the downstream secondary focus, which can be matched using the extraction line quadrupoles. The beam at the secondary focus was made parallel to the beam at the interaction point. The preferred value of R22=-0.5 from the interaction point to the secondary focus keeping a small spot size at the second focus has been obtained by re-matching the extraction line quadrupoles.

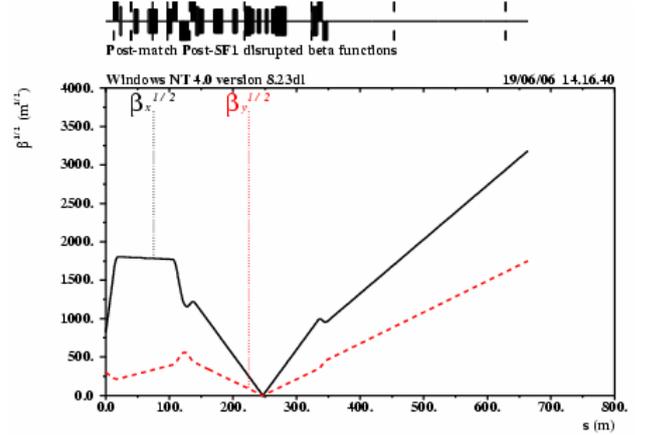

Figure 1: The matched linear optics for the 500 GeV CM extraction line, using the optimised NbTi final doublet layout.

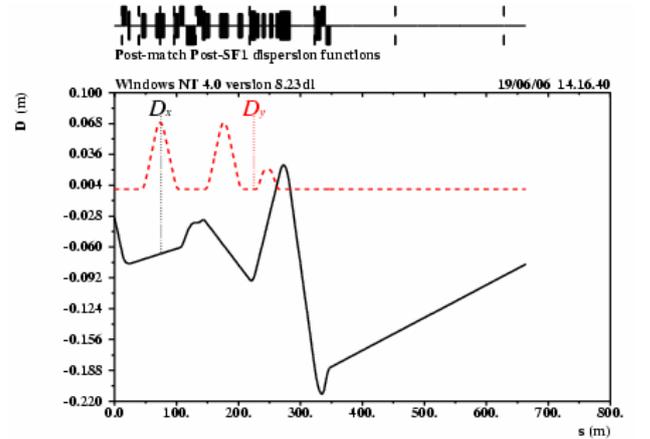

Figure 2: The matched dispersion functions for the 500 GeV CM extraction line, using the optimised NbTi final doublet layout.

The matched linear optics is shown in figure 1 and the dispersion for the extraction line is shown in figure 2. The optical performance is comparable to the baseline design, and satisfies the extraction line design criteria. The beam transport properties, charged beam power losses and the resulting interaction region backgrounds are currently under study. For example, the beam size at the first extraction line quadrupole magnet and the power losses on the first extraction line collimator are determined by the final doublet properties, and need to be carefully studied. Preliminary studies show that power losses on the collimator placed in front of the first extraction septum quadrupole have increased, indicating the need for further optimisation. The lengths of the extraction line magnets can also be reduced for the operation at 500 GeV CM and improved sextupole configurations can be considered to reduce their physical sizes. These optimisation studies will be done to reduce the losses on the magnets. The geometry of the extraction line needs to be fixed for the future upgrades and the next step will be to consider this for the upgrade to 1 TeV using the new final doublets.

## CONCLUSIONS

In this paper we present improved final doublet and extraction line layouts for the 2mrad crossing angle scheme of the ILC. The baseline design contains downstream collimation and diagnostic chicanes, but exhibits unsatisfactory beam power losses for some of the ILC parameters. We consider current and proposed superconducting magnet technologies to redesign the final doublet region, and present corresponding extraction line optics. The new extraction line is specifically optimised for the 500 GeV CM machine.

The overall optimisation of the 2mrad extraction line is ongoing, and requires considerable design studies. One area of difficulty is the realistic engineering design of the specialised extraction line magnets. These magnets need large fields and large apertures in order to handle the outgoing disrupted beam and beamstrahlung photons; but they need to be small enough to leave zero-field room for the incoming beam about 20cm beyond the photons. Efforts to find feasible extraction magnet solutions are continuing. A further magnet issue is the large external size of the final doublet magnets, which become large in the current layouts. Additional ongoing work includes a study of extraction line beam transport, validation of the diagnostic performance and computation of the detector backgrounds arising from post-IP particle and photon loss.